\documentclass[12pt]{article}
\usepackage{graphicx}
\newcommand{\lsim}{\,\lower.5ex\hbox{$\stackrel{\textstyle <}{\sim}$}\,}
\begin{document}
\title{In Search of the $f_0$ (or ``$\sigma$'') Meson: New data on
$\pi^0\pi^0$ Production by $\pi^-$ and $K^-$ on
Hydrogen\footnote{Contribution to the International Workshop on Chiral
Fluctuations in Hadronic Matter, Orsay, France, September 2001.}} 

\author{B.M.K.~Nefkens\footnote{\texttt{nefkens@physics.ucla.edu}}, 
        S.~Prakhov, and A.~Starostin\\
        \emph{UCLA, Los Angeles, CA 90095--1547, USA},\\
        for the Crystal Ball Collaboration
\footnote{Supported in part by US DOE, NSF, NSERC, RMS and VS.}
\footnote{The Crystal Ball Collaboration: M.~Clajus, A.~Maru\u{s}i\'c,
S.~McDonald, B.M.K.~Nefkens, N.~Phaisangittisakul, S.~Prakhov,
J.W.~Price, A.~Starostin, and W.B.~Tippens, \emph{UCLA}; D.~Isenhower
and M.~Sadler, \emph{ACU}; C.~Allgower and H.~Spinka, \emph{ANL};
J.~Comfort, K.~Craig, and T.~Ramirez, \emph{ASU}; T.~Kycia, \emph{BNL};
J.~Peterson, \emph{UCo}; W.~Briscoe and A.~Shafi, \emph{GWU};
H.M.~Staudenmaier, \emph{UKa}; D.M.~Manley and J.~Olmsted, \emph{KSU};
D.~Peaslee, \emph{UMd}; V.~Bekrenev, A.~Koulbardis, N.~Kozlenko,
S.~Kruglov, and I.~Lopatin, \emph{PNPI}; G.M.~Huber, G.J.~Lolos, and
Z.~Papandreou, \emph{UReg}; I.~Slaus and I.~Supek, \emph{RBI};
D.~Grosnick, D.~Koetke, R.~Manweiler, and S.~Stanislaus, \emph{ValU}.}}
\date{January 14, 2002}
\maketitle

\begin{abstract} 
We present preliminary results on the cross sections and Dalitz-plot
densities for the process $\pi^-p\to\pi^0\pi^0 n$ from threshold to
$p_\pi=750$ MeV/$c$ as well as for $K^-p\to\pi^0\pi^0\Lambda$ and
$K^-p\to\pi^0\pi^0\Sigma^0$ at $p_K = 520$ to 750 MeV/$c$.  We have
found that $\sigma_{tot}(\pi^-p\to\pi^0\pi^0n) \simeq
2\sigma_{tot}(K^-p\to\pi^0\pi^0\Lambda)$.  The $\pi^0\pi^0n$ Dalitz
plots are very nonuniform, expecially for the higher $p_{\pi}$, with a
high concentration of events on an ``island" around $m(\pi n)\simeq$
1.2 GeV and $\Gamma\simeq$ 0.1 GeV peaking at high $\pi^0\pi^0$
invariant mass.  This is indicative of the dominant role of the
$\Delta^0(1234)\frac{3}{2}^{+}$ resonance in the final state. 

The $\pi^0\pi^0 \Lambda$ Dalitz plots are strikingly similar to the
ones for $\pi^0 \pi^0 n$ except that the island is concentrated at
$m(\pi\Lambda)\simeq$ 1.38 GeV and has a narrower width,
$\Gamma\simeq0.05$ GeV.  This indicates the dominant role of the
$\Sigma^0(1385)\frac{3}{2}^+$ resonance.  The similarity in the Dalitz
plots and the proportionality of the total cross sections are an
impressive testimony of the applicability of broken $SU(3)$ flavor
symmetry to reaction dynamics.  We have measured
$\sigma_{tot}(K^-p\to\pi^0\pi^0\Lambda) \simeq
6\sigma_{tot}(K^-p\to\pi^0\pi^0\Sigma^0)$ and observed that the Dalitz
plots for 
these processes are very different. The Dalitz plots for
$\pi^0\pi^0\Sigma^0$ show some enhancement at low
$\pi^0\pi^0$ invariant mass, and there is good indication for the
$\Lambda(1405)$ intermediate state but there is no island; at the
highest $p_K$, there is some evidence for the $\Lambda(1520)$
intermediate state.  The above features of $\pi^0\pi^0$ production by
$\pi^-$ and $K^-$ can all be understood if $f_0$ production is small.
\end{abstract}

\section{INTRODUCTION}

The $f_0(400-1200)$ state, $I^G(J^{PC})= 0^+(0^{++})$, is the chiral
partner of the $\pi$ meson.  The original symbol for the $f_0$  is the
$\sigma$.  To minimize mixups when discussing the production of the
$\sigma$ and the $\Sigma$ in the same reaction we will use the symbol
$f_0$.  This is the notation employed by the Particle Data
Group~\cite{PDG01}.  The $f_0$ is a broad and not well-determined 
state; recommended in \cite{Ish00} are a mass around 500 MeV and
width in the vicinity of 500 MeV also.  The large width ensures
that the 
$f_0$ decays inside the nucleus in which it is produced.  The $f_0$ is
a prime candidate to investigate hadron medium modifications which
have been the subject of much theoretical discussion as witnessed in
this workshop.  The medium modifications include a change in the mass
and width when the hadron is embedded in nuclear matter of
sufficiently high density.  The $f_0$ decays $\sim 100\%$ into two
pions.  The $\pi^0 \pi^0$ decay mode is particularly attractive as two
$\pi^0$'s must be in an even $I$ and $J$ state.  This avoids the
troublesome $I$=1 $\rho$ contribution which is  present in experiments
with the $\pi^+\pi^-$ final state.  Furthermore, $\pi^0$'s can be
measured by the 2$\gamma$ decay at all kinetic energies down to
$T_\pi=0$ MeV.  This leads to a high and uniform acceptance for the
$\pi^0 \pi^0$ system at all invariant masses down to 270 MeV.

We present preliminary  results on $\pi^0 \pi^0$ production in the
process $\pi^- p \to \pi^0 \pi^0 n$  from threshold to $p_\pi = 750$
MeV/$c$, and for $K^- p \to \pi^0 \pi^0 \Lambda$ and $K^- p \to \pi^0
\pi^0 \Sigma^0$ at $p_K = 520 - 750$ MeV/$c$.  These three reactions,
together with flavor symmetry, are very useful for probing the
reaction mechanism responsible for $2 \pi^0$ production.  It will help
in the investigation of the unexpected claim~\cite{Bon96,Bon00} that
medium modification of the $f_0$ meson was sighted in $\pi^+\pi^-$
production by $\pi^+$ at $p_\pi \simeq$ 400 MeV/$c$ on ordinary nuclei
of standard nuclear density. 

\section{THEORETICAL CONSIDERATIONS}

At the beam momenta of this experiment the $2\pi^0$ production process,
\begin{equation}
\pi^- p \to \pi^0 \pi^0 n,
\end{equation}
is expected to be dominated by  $s$-channel amplitudes leading to
$N^*$ formation.  The subsequent $N^*$ decay can occur in two
different ways:
\begin{itemize}
\item [a)] by the decay of an intermediate state meson, the $f_0$: 
\begin{equation}
\pi^-p\to N^*\to f_0n\mathrm{\ \ followed\ by\ \ }f_0\to\pi^0\pi^0;
\end{equation}
\item [b)] via the decay of a second intermediate state baryon resonance,
the $\Delta$:
\begin{equation}
\pi^- p\to N^*\to \pi^0 \Delta^0\mathrm{\ \  followed\  by\ \ }  \Delta^0
\to \pi^0n.
\end{equation}
\end{itemize}
The pole and contact terms are small and may be ignored.  Reactions
(2) and (3) are interwoven with one another by final state
interactions which are energy dependent.  The $\pi^0-\pi^0$
interaction is dominated by $s$-wave scattering controlled by the 
$\delta_0^0$ phase.  The cross section for $\pi^0-\pi^0$ scattering
increases monotonically  from threshold to the peak of the
$f_0$, presumably around 500 MeV.  $\pi N$ scattering reaches a huge
peak when $m(\pi N)$ = 1232 MeV which corresponds to $p
_{\pi}$ = 227 MeV/$c$ in the c.m.  At low energy we expect the $\pi^0  -
\pi^0$ scattering to be bigger than $\pi^0 - n$ scattering, and the
reverse at higher energies.

The final state of the $\pi^- p \to \pi^0 \pi^0 n$ reaction will be
described using a Dalitz plot in which the vertical axis is the
invariant mass squared of the $\pi^0 \pi^0$ system, $m^2 (\pi^0
\pi^0)$, and the horizontal axis is the invariant mass squared of the
$\pi^0 n$ system, $m^2 (\pi^0n)$.  The final state features two
identical $\pi^0$'s.  Thus,  if the process is $\pi^- p \to \pi^0_1
\Delta^0$ followed by $\Delta^0 \to \pi^0_2 n$ we don't know which
$\pi^0$ is $\pi^0_1$ or $\pi^0_2$.  This complication can be handled
by making two entries in the Dalitz plot, recording both the
$m^2(\pi^0_1 n)$ and $m^2(\pi^0_2n)$ options.  All our Dalitz plots
have $2\pi^0$ and are handled this way. 

Figure~1a shows a Monte Carlo, MC, generated Dalitz plot, DP, for
$\pi^-p\to f_0n\to\pi^0\pi^0n$ for $m(f_0)=0.4$ GeV and
$\Gamma(f_0)=0.1$ GeV with a Breit-Wigner shape at $p_\pi=0.75$
GeV/$c$. 
\begin{figure}
\includegraphics[width=\textwidth]{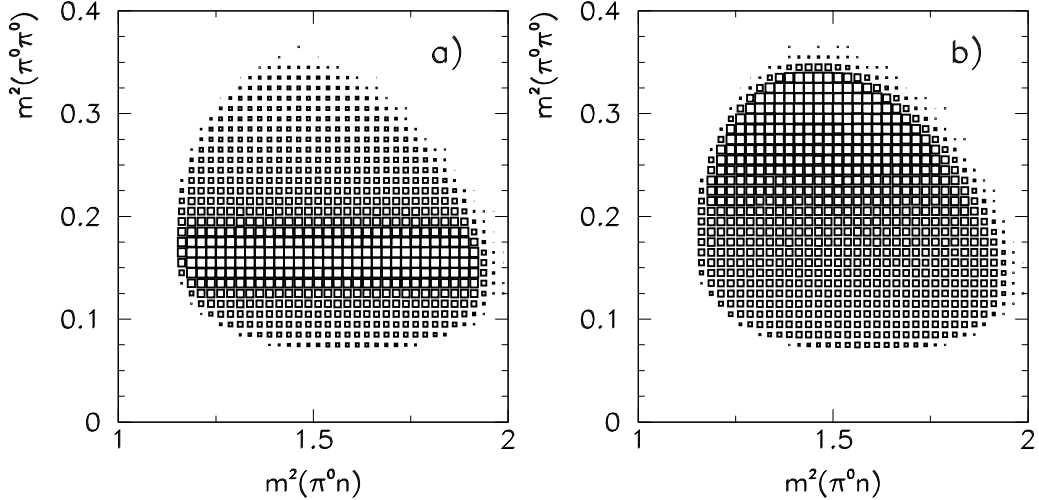}
\caption{\label{fig1}Monte Carlo generated Dalitz plots for $\pi^-p\to
f_0n\to\pi^0\pi^0n$ at $p_\pi=0.75\,\mathrm{GeV}/c$.  a) $m(f_0)=0.4$
GeV, $\Gamma(f_0)=0.10$ GeV.  b) $m(f_0)=0.75$ GeV,
$\Gamma(f_0)=0.40$ GeV.}  
\end{figure}
This
DP is characterized by a strong horizontal band around
$m^2(\pi^0\pi^0)=(0.4\,\mathrm{GeV})^2$, which has a uniform density
along the 
$m^2(\pi^0n)$ axis. The uniform density is a typical feature of the
$s$-wave decay of the $f_0$ state without interference.

Figure~1b shows another MC DP, for $m(f_0)=0.7$ GeV,
$\Gamma(f_0)=0.4$ GeV, also with a Breit-Wigner shape at $p_\pi=0.75$
GeV/$c$. 
There is a broad horizontal band which is uniform in horizontal
slices.  The density increases with increasing $m^2 (\pi^0\pi^0)$, it
is the consequence of choosing $m (f_0 ) =$\ 0.7 GeV.

The expected DP for pure $\Delta (1232)$ production, Eq.~3, is shown in
Fig.~2.  
\begin{figure}
\includegraphics[width=\textwidth]{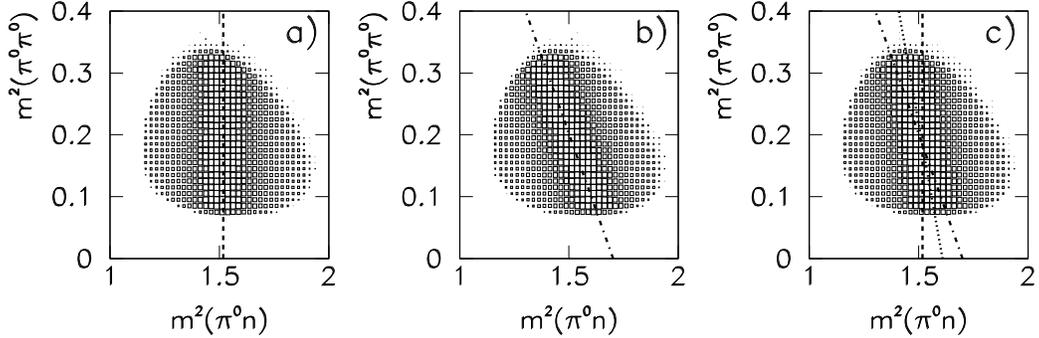}
\caption{\label{fig2}Monte Carlo generated Dalitz plots for $\pi^-p\to
\pi^0_1\Delta^0\to\pi^0_1\pi^0_2n$ at $p_\pi=0.73\,\mathrm{GeV}/c$.
a) $m(\pi n)=m(\pi^0_2n) = m(\Delta^0)$, that is when using the
``right'' $\pi^0$.
b) $m(\pi n)=m(\pi^0_1n)$, for the alternate $\pi^0$.
c) The sum of the above two plots.
The dashed line in a) and c) is the pole position of the $\Delta$ at
1.21 GeV.  The dashed-dotted line in b) and c) is the reflection of
the $\Delta$ pole position.  The dotted line in c) is the line of
symmetry of the DP.}  
\end{figure}
For the MC we used a Breit-Wigner shape of the $\Delta$ with
$m(\Delta)=\ $1.2 GeV and $\Gamma(\Delta)=\ $0.1 GeV.  For simplicity,
we used an isotropic decay of the $\Delta$.  Since the
MC calculation $\lq\lq$knows" which of the two $\pi^0$'s belongs to
the $\Delta$ we can investigate the consequences of our way of
handling the two identical $\pi^0$'s.  For the correct $\pi^0$ choice,
shown in Fig.~2a, one finds the expected, uniform \emph{vertical} band.  The
wrong $\pi^0$ choice results in a slightly slanted vertical band, see
Fig.~2b.  This wrong band may be constructed as the reflection of the
correct band on the symmetry line that characterizes every DP in which
two particles are identical.  The symmetry line is a straight line
connection $m^2_\mathit{max}(\pi^0\pi^0)$ to
$m^2_\mathit{min}(\pi^0\pi^0)$.  For every event one entry will be on
the left, the 
other 
on the right side of the symmetry line at equal distance.   The sum of
Figs.~2a and 2b is shown in Fig.~2c.  It is a broadened, somewhat less
slanted band than in Fig.~2b.  It
is clear now that the projection of the Dalitz plot content on the
$m^2 (\pi^0n)$ axis does not show the correct mass and width
of the 
$\Delta!$   These may be obtained by unfolding the double entry
feature of the DP, which we will do using a Monte Carlo based 
technique. 

As the $\Delta$ decay is simulated with an isotropic distribution, the
bands in Figs.~2a and 2b are of uniform density.  Since the
production reaction is actually via $N^* \to \pi^0 \Delta^0$ the DP
density distribution must reflect the angular momentum involved in the
$N^*$ and $\Delta$ decay. The intermediate state
$N^*$ resonance can be the Roper, $N(1440) \frac{1}{2}^+$, which
requires an $\ell = 1$ transition to the $\pi \Delta$ state.  It could
also be the $N(1520)\frac{3}{2}^-$ which implies $\ell=0$ and even the
$N(1535)\frac{1}{2}^-$ which needs $\ell=2$. 

The energy dependence of the total cross section for $2 \pi^0$
production should reflect the production rate of the different $N^*$
resonances.  Thus, we expect that $\sigma_{tot} (\pi^- p \to \pi^0 \pi^0
n)$  will show a steady increase from threshold to our maximum beam
momentum of 750 MeV/$c$.  

The full QCD Lagrangian, ${\cal L}_{QCD}$, may be divided into two parts,
\begin{equation}
{\cal L}_{QCD} = {\cal L}_0 + {\cal L}_m.
\end{equation}
The first part, ${\cal L}_0$, depends only on the quark and gluon
fields, but not on the quark masses; the second part, ${\cal L}_m$,
depends on the magnitude of the current quark masses.
In the limit of massless quarks ${\cal L}_{QCD}$ is equal to ${\cal
L}_0$ which is the same for all
quarks.  This is the famous (quark) flavor symmetry of QCD.  It
implies that in the limit of massless quarks the interaction of two
systems of particles which differ only by the replacement of a
$d$-quark by a $s$-quark but otherwise have the same $SU(3)$ flavor
symmetry are described by the same ${\cal L}_0$ and thus have
identical strength.  This means that both reactions will have the same
cross section, identical DP's, polarizations, and so forth.  In the
real world the quarks have masses and flavor symmetry is broken.  The
breaking is given by the mass term ${\cal L}_m$ of the full ${\cal
L}_{QCD}$. Limiting ourselves to the three light quarks it is simply
\begin{equation}
{\cal L}_m = - {\bar{u}}m_u u -{\bar{d}} m_d d - {\bar{s} } m_s s
\end{equation}
${\cal L}_m$ acts mainly as a correction to ${\cal L}_0$.  In the
following we will make 
the quark-model assumption that light mesons are $q\bar{q}$ and light
baryons $qqq$ states.  Recently a stunning case of the applicability of
(broken) $SU(3)$ flavor symmetry was observed.  It is the (semi)
quantitative agreement between the characteristics of threshold $\eta$
production,  $\sigma_{tot}$, $d\sigma/dE$, $d\sigma/d\Omega$, etc.\ of
the flavor-symmetric reactions $\pi^-p\to\eta n$ and
$K^-p\to\eta\Lambda$ \cite{Nef01,Sta01}. 

Now we would like to compare $2 \pi^0$ production via the $f_0$-meson
intermediate state:
\begin{eqnarray}
&\pi^-p \to N^*       \to f_0n        \to \pi^0\pi^0n,\\
&K^-p   \to \Lambda^* \to f_0\Lambda  \to \pi^0\pi^0\Lambda,\\
&K^-p   \to \Sigma^*  \to f_0\Sigma^0 \to \pi^0\pi^0\Sigma^0.
\end{eqnarray}
The incident $\pi^-$ and $K^-$ belong to the same $SU(3)$ pseudoscalar
meson octet.  The final state $n, \Lambda$ and $\Sigma$ belong to the
same $\frac{1}{2}^+$ baryon octet.  The allowed $N^*$, $\Lambda^*$,
and $\Sigma^*$ intermediate states also belong to the same $SU(3)$
baryon octets, see table~\ref{tab:intstates}.   Thus, if $2\pi^0$
production would occur dominantly via the $f_0$ intermediate state, we
would expect that the three reactions in Eqs. 6--8 have similar DP
density distributions and comparable cross sections.  Specifically,
without correcting for the different SU(2) and SU(3) C-G coefficients
and phase spaces, we have 
\begin{equation}
\label{eq:flavsymf0}
\sigma_{tot}(\pi^-p\to\pi^0\pi^0n) =
\sigma_{tot}(K^-p\to\pi^0\pi^0\Lambda) = 
\sigma_{tot}(K^-p\to\pi^0\pi^0\Sigma^0).
\end{equation}
\begin{table}
\caption{\label{tab:intstates}The Flavor Symmetric $N^*$, $\Lambda^*$,
and $\Sigma^*$ intermediate states in $2\pi^0$ production.  States in
the same row are flavor-symmetric with each other.}
\begin{center}
\begin{tabular}{c@{$\ \leftrightarrow\ $}c@{$\ \leftrightarrow\ $}c}
\hline\hline
\multicolumn{1}{c}{$N^*$} & 
\multicolumn{1}{c}{$\Lambda^*$} & 
\multicolumn{1}{c}{$\Sigma^*$} \\
\hline
$N(1440)\frac{1}{2}^+$ & 
$\Lambda(1600)\frac{1}{2}^+$ &
$\Sigma(1660)\frac{1}{2}^+$ \\ 
$N(1535)\frac{1}{2}^-$ &
$\Lambda(1670)\frac{1}{2}^-$ & 
$\Sigma(1620)\frac{1}{2}^-$ \\
$N(1520)\frac{3}{2}^-$ &
$\Lambda(1690)\frac{3}{2}^-$ & 
$\Sigma(1670)\frac{3}{2}^-$ \\
\hline \hline
\end{tabular}
\end{center}
\end{table}

On the other hand, $2\pi^0$ production may occur by sequential
baryon resonance deexcitation:
\begin{eqnarray}
&\pi^-p \to  N^*       \to \pi^0\Delta^0(1232)  \to \pi^0\pi^0n,
\label{eq:pip2pipin} \\
&K^- p  \to  \Lambda^* \to \pi^0\Sigma^0(1385)  \to \pi^0\pi^0\Lambda^0,
\label{eq:kp2pipilam}\\
&K^- p  \to  \Sigma^*  \to \pi^0\Lambda(1405/1520)  \to \pi^0\pi^0\Sigma^0.
\label{eq:kp2pipisig}
\end{eqnarray}
The initial and the final states are flavor symmetric.  The $\Delta
(1232)\frac{3}{2}^+$ and $\Sigma(1385)\frac{3}{2}^+$ in
Eqs.~\ref{eq:pip2pipin} and \ref{eq:kp2pipilam} belong to the same
$SU(3)$  decuplet thus they are flavor symmetric.  We predict that the
DP's and cross sections will be similar for $\pi^- p \to \pi^0 \pi^0
n$ and $K^- p \to \pi^0 \pi^0 \Lambda$.   $SU(3)$ breaking may be
accounted for by comparing at incident beam momenta such that both
reactions have the same $m_\mathit{max}(\pi^0 \pi^0)$.  Also, we predict
that the $\Delta (1232) $ band in the DP should have three times the width
of the $\Sigma(1385)$band because
$\Gamma(\Delta^0)\simeq3\Gamma\{\Sigma^0(1385)\}$.  We also predict that  
\begin{equation}
\label{eq:flavsymtheo}
\sigma_{tot}(K^-p\to\pi^0\pi^0n) = 
(2\pm0.5)\sigma_{tot}(K^-p\to\pi^0\pi^0\Lambda),
\end{equation}
where $(2\pm0.5)$ is the product of the $SU(2)$ and $SU(3)$
Clebsch-Gordan coefficients and a 
phase space correction factor.   The $\Lambda (1405) \frac{1}{2}^-$ and
$\Lambda(1520) \frac{3}{2}^-$  are $SU(3)$ singlet states.  There is
\emph{no} flavor symmetry between Eqs.~\ref{eq:kp2pipilam} and
\ref{eq:kp2pipisig}, and we predict that the $\pi^0 \pi^0 \Lambda$ and
$\pi^0 \pi^0 \Sigma^0$  DP's will be different and also that 
\begin{equation}
  \sigma_{tot} (K^- p \to \pi^0 \pi^0 \Lambda)   \not= \sigma_{tot}(K^- p
\to \pi^0 \pi^0 \Sigma^0).
\end{equation}

\section{Experiment}
2$\pi^0$ production has been measured at the AGS at Brookhaven
National Laboratory in the C6 line using a range of separated $\pi^-$
and $K^-$ beams up to 750 MeV/$c$.  The uncertainty in the absolute
value of the incident beam momentum is $<1\%$; $\Delta p/p$ is
typically $3.5\%$.  The detector was the Crystal Ball (CB) 
multiphoton spectrometer, which consists of 672 separate NaI counters, 16
$X_0$ deep, it covers $93\%$ of the full $4\pi$ solid angle.  The
CB has good energy and angular resolutions:
$\sigma_E/E=1.7\protect \% /\{E(GeV)\}^{0.4} $and $\sigma_\theta =
2^{\circ} - 3^{\circ}$.  A liquid $H_2$ target is located in the hollow
center of the ball.  The target is surrounded by a plastic veto
counter system for triggering on neutral final states.  Details of the CB and
the analysis are given in Ref. \cite{Sta01,Cra01,Sta00}.  To measure
$\pi^- p \to \pi^0 \pi^0 n$ it is sufficient to detect the four gammas
from the $2 \pi^0$ decay.  They are part of the four-gamma-cluster
event sample. The
$\Lambda$ is detected by the $\pi^0$ from its $n \pi^0$ decay, thus
the $\pi^0 \pi^0 \Lambda$ final state is found in the six-cluster events.
The $\Sigma^0$ has one more gamma from $\Sigma^0 \to
\Lambda \gamma$ decay, thus the $\pi^0 \pi^0 \Sigma^0$ final state is
a seven-photon cluster event.

\section{The total Cross Sections for $2 \pi^0$  Production}
The preliminary results obtained in the Crystal Ball experiment for
$\sigma_{tot}(\pi^-p\to\pi^0\pi^0n)$ at 17 incident $\pi^-$ momenta,
and for $\sigma_{tot}(K^-p\to\pi^0\pi^0\Lambda)$ and
$\sigma_{tot}(K^-p\to\pi^0\pi^0\Sigma^0)$ at eight incident $K^-$
momenta are shown in Fig.~3. 
\begin{figure}
\includegraphics[width=\textwidth]{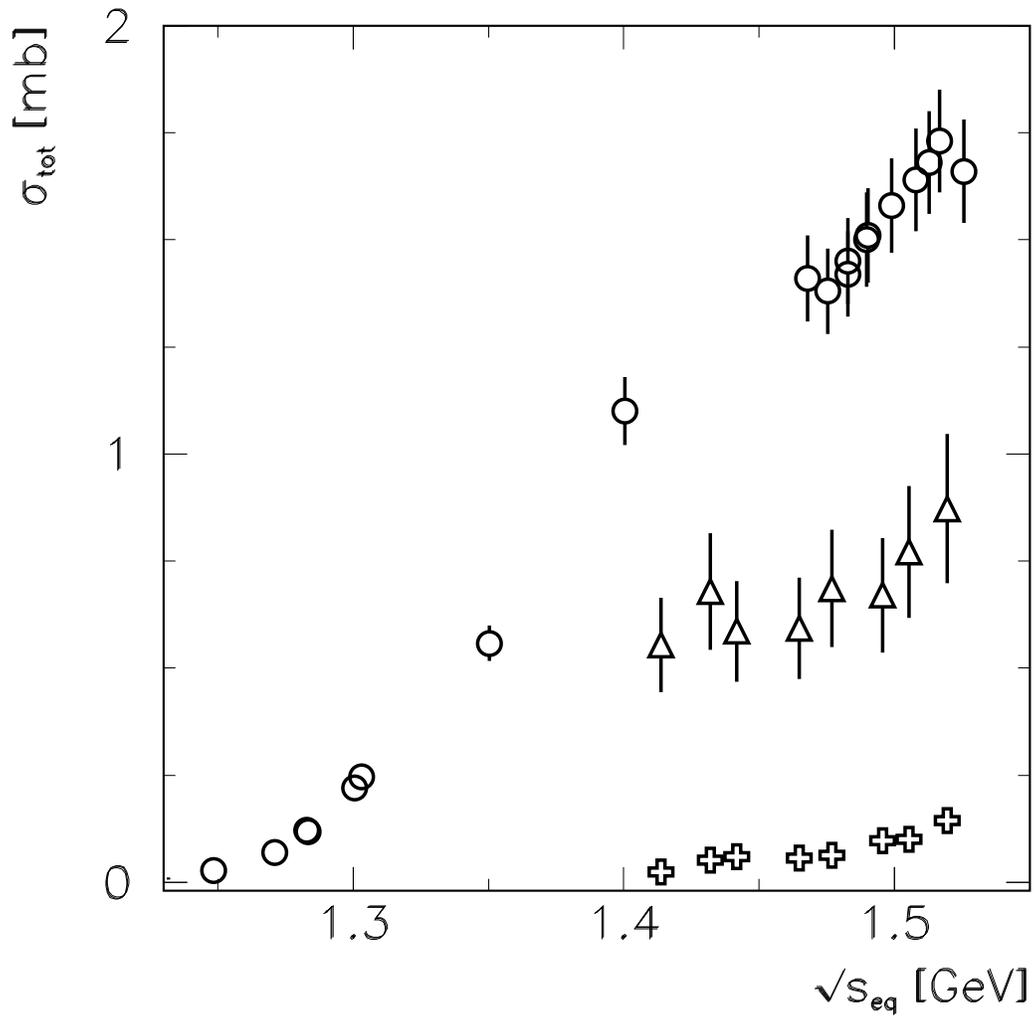}
\caption{\label{fig3}The total cross sections as functions of
$\sqrt{s_{eq}}$.  
Circles: $\sigma_{tot}(\pi^-p\to\pi^0\pi^0n)$.
Triangles: $\sigma_{tot}(K^-p\to\pi^0\pi^0\Lambda)$.
Crosses: $\sigma_{tot}(K^-p\to\pi^0\pi^0\Sigma^0)$.}  
\end{figure}
The
variable which we use on the abscissa is the \emph{equivalent total
energy} $\sqrt{s_{eq}}$, where $s_{eq}$ for incident pions is the
standard $s$.  For incident kaons we define  $\sqrt{s_{eq}} \equiv
\sqrt{s} - (m_s - m_d)$.  This is one of several ways for
incorporating a correction for the $s$-$d$ quark mass difference.  We
will use $m_s - m_d = 157$ MeV obtained from the systematics of the
baryon-multiplet masses \cite{Nef95}. Experimentally we have found that
\begin{equation}
\label{eq:flavsymexp}
\sigma_{tot}(\pi^-p\to\pi^0\pi^0n) = 
(2\pm0.4)\sigma_{tot}(K^-p\to\pi^0\pi^0\Lambda)
\end{equation}
in the $\sqrt{s_{eq}} $ span of 1.41 to 1.53 GeV.  The
difference in the phase space of the $2 \pi^0 n$ and $2 \pi^0 \Lambda$
final states is small and can be ignored here.  Another way to correct
for the $s$-$d$ quark mass difference is by comparing at those incident beam
momenta for which $m_{max}(\pi^0\pi^0)$ is the same.  In the case
at hand it gives very similar results to Eq.~\ref{eq:flavsymexp}. 

The agreement between the prediction presented in
Eq.~\ref{eq:flavsymtheo} and the data given in Eq.~\ref{eq:flavsymexp}
demonstrates the applicability of flavor symmetry in $2\pi^0$
production.  This is a remarkable result as we are comparing the
dynamics of three-body final-state reactions. 

Figure~3 illustrates that $\sigma_{tot}(K^-p\to\pi^0\pi^0\Sigma^0)$ is
smaller than $\sigma_{tot}(K^-p\to\pi^0\pi^0\Lambda)$ by a
factor of 6 and more at the same $\sqrt{s}$.  If these 
reactions would occur via $f_0$  production in the intermediate state
their cross sections should be comparable, see Eq.~\ref{eq:flavsymf0}.
Our results 
are consistent with $2 \pi^0$ production by $K^-$ occuring by
sequential baryon-resonance decay,
Eqs.~\ref{eq:pip2pipin}--\ref{eq:kp2pipisig}.  After correcting for
the 
$\Lambda -  \Sigma^0$ mass difference our data give for the corrected
total cross sections $\sigma^c_{tot}$
\begin{equation}
\label{eq:f0percent}
\sigma^c_{tot}(K^-p\to\pi^0\pi^0\Lambda) =
(5\pm1)\sigma^c_{tot}(K^-p\to\pi^0\pi^0\Sigma^0), 
\end{equation}
using either the method of comparing at the $\sqrt{s}$ after correcting
for the difference in the $2 \pi^0 \Lambda$ and $2 \pi^0 \Sigma$ final
states, or by comparing at the same $m(\pi\pi)$, the latter is shown
in Fig.~4.  
\begin{figure}
\includegraphics[width=\textwidth]{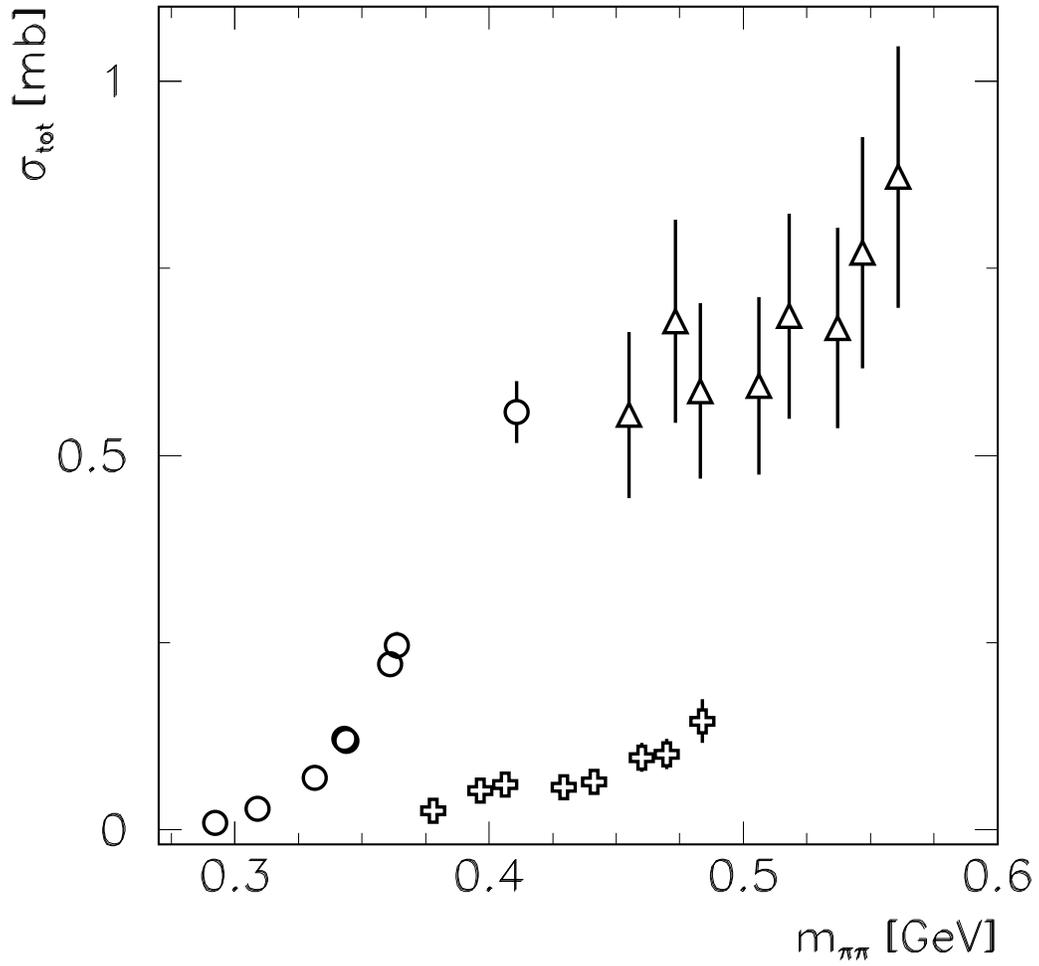}
\caption{\label{fig4}Same as Fig.~3, with
$m_{\mathit{max}}(\pi^0\pi^0)$ plotted on the $x$-axis.}  
\end{figure}

We conclude from our data on $2 \pi^0$production by $K^- $ that the
$f_0$ does not play a major role.  From Eq.~\ref{eq:f0percent} we may
conclude that 
there is at most  20\% $f_0$ production.  Analysis of the Dalitz Plots
in the following sections shows it is actually even smaller.  To
investigate the possible onset of chiral  restoration it would be of
interest to measure $\pi^0 \pi^0 \Lambda$ and $\pi^0 \pi^0
\Sigma^0$ production by $K^-$ on complex nuclei. 

\section{The $\pi^0\pi^0n$ Dalitz Plots}
The DP's for $\pi^- p \to \pi^0 \pi^0 n$ at 8 incident
$p_{\pi}$ are shown in Fig.~5.  
\begin{figure}
\includegraphics[width=\textwidth]{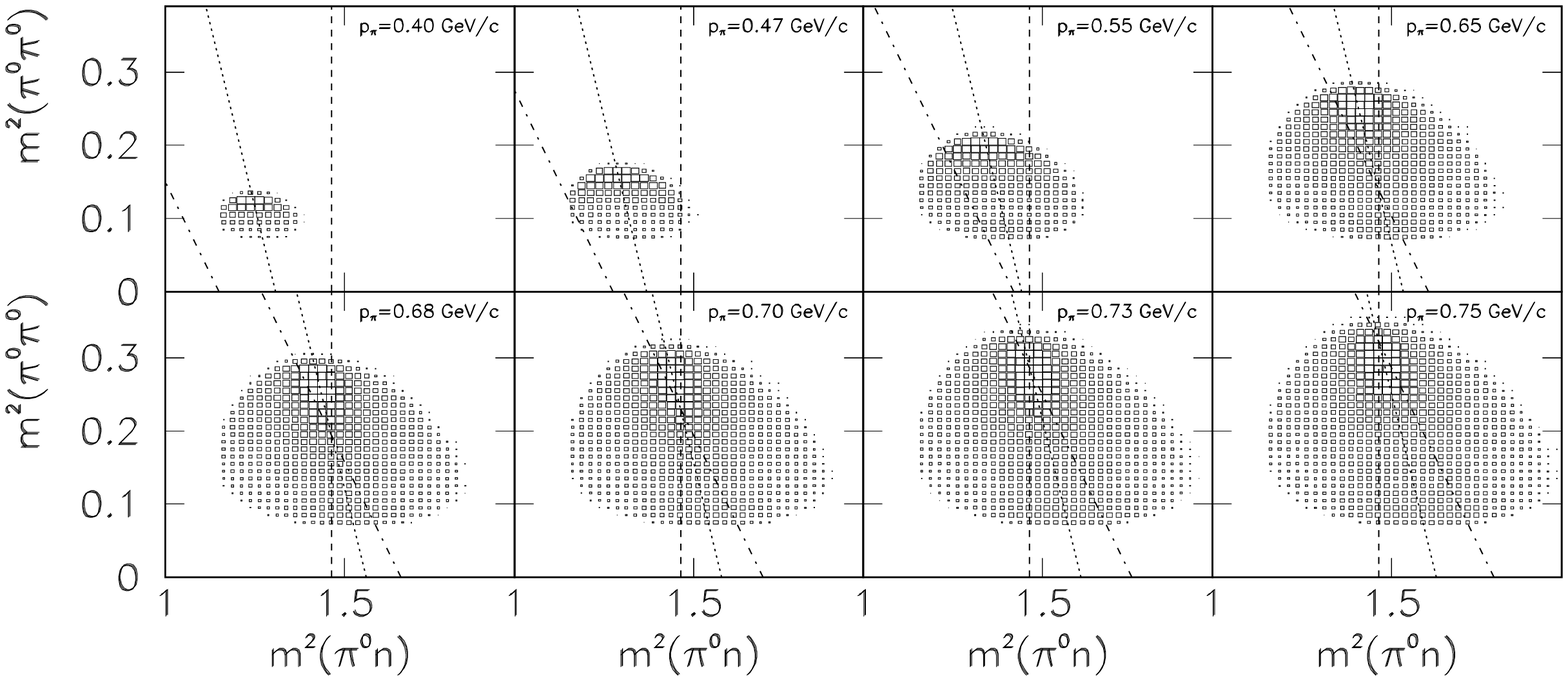}
\caption{\label{fig5}Dalitz plots measured for $\pi^-p\to\pi^0_1\pi^0_2n$.
Dotted line: line of symmetry.  There are two entries for every event,
namely $m^2_1=m^2(\pi^0_1n)$, and $m^2_2=m^2(\pi^0_2n)$.
Dashed line: predicted position for $m^2(\pi^0n)=m^2(\Delta^0)$.
Dashed dotted line: $m^2(\pi n)$ for the alternate $\pi^0$ choice,
using $m(\Delta)=1.21$ GeV (pole position).}  
\end{figure}
For convenience in the analysis we
have divided the  data into 2 sets of $p_\pi$, a high and a low $p_\pi$.

The high $p_\pi$ set covers $p_\pi$ from 650 to 750 MeV/$c$ when
$m(\pi\pi)$ extends from threshold at 0.27 GeV to 0.59 GeV, while $m(\pi
n)$ goes from threshold at 1.08 GeV to 1.40 GeV, thus, the
$\Delta(1232)$ is near the center of the DP. We have measured a total
of nine DP's in 
this range, but only five are shown in Fig.~5. All DP's of the high
$p_\pi$ set have the same gross features: the density is very
non-uniform along both the horizontal and vertical axes.  There is a
high density region called an ``island" located at
$m^2(\pi^0n)\simeq(1.2\,\mathrm{GeV})^2$ in the upper part of the DP's
near the 
maximum allowed $m^2(\pi^0\pi^0)$ value.  There is also a minor enhancement,
it is called the ``reef" at $m^2(\pi^0n)\simeq(1.3\,\mathrm{GeV})^2$ in the
lower part of the DP's near the $m(\pi^0\pi^0)$ threshold. There is no
evidence for a horizontal band of uniform density which  would reveal
dominant $2\pi^0$ production via the $f_0$.  The island 
and reef together form a slightly slanted vertical band along the line
of symmetry of each DP which is shown as a dotted line in each of the
DP's of Fig.~5.  If $2\pi^0$ production takes place exclusively via
the sequence 
\begin{equation}
\pi^- p \to N^* \to \pi^0_1 \Delta^0 \to \pi^0_1 \pi^0_2 n,
\end{equation}
the centroid of the $m^2 (\pi^0_2 n)$ distribution is the centroid of the
$\Delta^0$ distribution which is indicated by a dashed
\emph{vertical} 
line.  Since it is not known which $\pi^0$ is $\pi^0_1$ or $\pi^0_2$
we have plotted the $m^2 (\pi^0_1n)$ distribution as well, its maximum is shown
by the slanted dashed dotted line.  Note that the $m^2(\pi^0_1n)$
distribution is 
the reflection of $m^2 (\pi^0_2 n)$ on the line of symmetry.  The
structrue of the DPs indicates that the reaction chain of Eq.~10 is
the dominant one.  The difference between Figs.~2 and 5 is likely the
forward-backward asymmetry of $\Delta$ decay which is seen in Fig.~5
as the island and the reef in each DP.  The origins of this asymmetry
are several, they include the angular momentum changes in the two
decays in Eq.~10.  This is the subject of a separate analysis.

For a simple qualitative analysis we have
divided the DP's at $p_\pi =$ 750, 730 and 650 MeV/$c$ into an upper
half which contains the island, and a bottom half which includes the
reef.  We reproduced all six $m^2(\pi^0 n)$ distributions by a Monte
Carlo simulation in which we use the same value for the $\Delta$ mass,
namely 1.21
GeV and the same width of 0.10 GeV.  These values are the pole
position of the $\Delta(1232)$ resonance. 

The set of DP's at low incident pion momentum ($p_\pi \leq 470$ MeV/$c$)
covers $m(\pi^0\pi^0)$ from 270 to 415 MeV and $m(\pi^0n)$ from 1.07
to 1.21 GeV.  The latter implies that only the lower tail of the
$\Delta$ peak can contribute.  Qualitatively, there is a large
excess of events over phase space in the upper half, the island is
broadened and the reef appears to be gone. The quantitative analysis of
these DP's is not yet completed.  

\section{The Dalitz Plots for $K^- p \to \pi^0\pi^0\Lambda$}
Figure~6 shows the DP's of our $K^- p \to \pi^0 \pi^0 \Lambda$ data for
all eight $K^-$ beam momenta.  
\begin{figure}
\includegraphics[width=\textwidth]{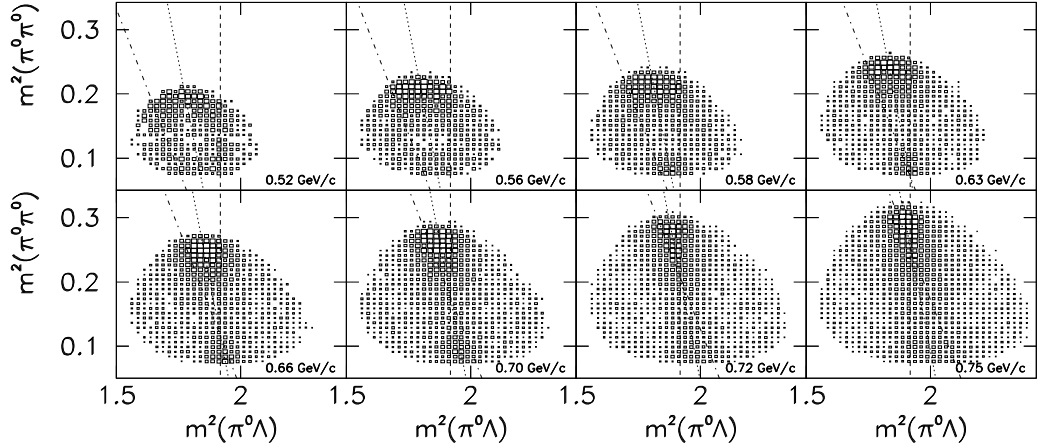}
\caption{\label{fig6}Same as Fig.~5 for the process
$K^-p\to\pi^0\pi^0\Lambda$. 
Dashed line: predicted position for $m^2(\pi^0\Lambda)=m^2(\Sigma^*)$,
with $m(\Sigma^*)=1.385$ GeV.}  
\end{figure}
The similarity in the density
distribution of  the $\pi^- p \to \pi^0 \pi^0 n$ DP's (see Fig.~5), at
similar $\sqrt{s_{eq}}$  (which is this case also means similar
$m(\pi^0\pi^0)$) is stunning.  It is a major  triumph for flavor
symmetry to relate the dynamics of two three-body final-state
reactions. The chief difference between Figs.~5 and 6 is in the
location of the  center and in the width of the islands and reefs.
The differences originate in the mass and width differences of the
$\Delta$ and $\Sigma^*$, namely $m(\Delta)=1.21$ GeV and
$\Gamma(\Delta)=0.10$ GeV, while $m(\Sigma^*)=1.38$ GeV and
$\Gamma(\Sigma^*)=0.05$ GeV. 

Seven of the eight DP's belong to the high-incident-beam-momentum
set, introduced in the previous section.  This means each DP has an
island and a reef which is determined by the mass and width of the
$\Sigma(1385)$ intermediate state resonance.  The value of $m(\pi^0\pi^0)$
covered in this part of the experiment extends from 0.27 to 0.46 GeV. 

\section{Dalitz Plots for $K^-p \to \pi^0\pi^0\Sigma^0$}
The DP's for the eight CB measurements of $K^- p \to \pi^0 \pi^0
\Sigma^0$ are shown in Fig.~7.  
\begin{figure}
\includegraphics[width=\textwidth]{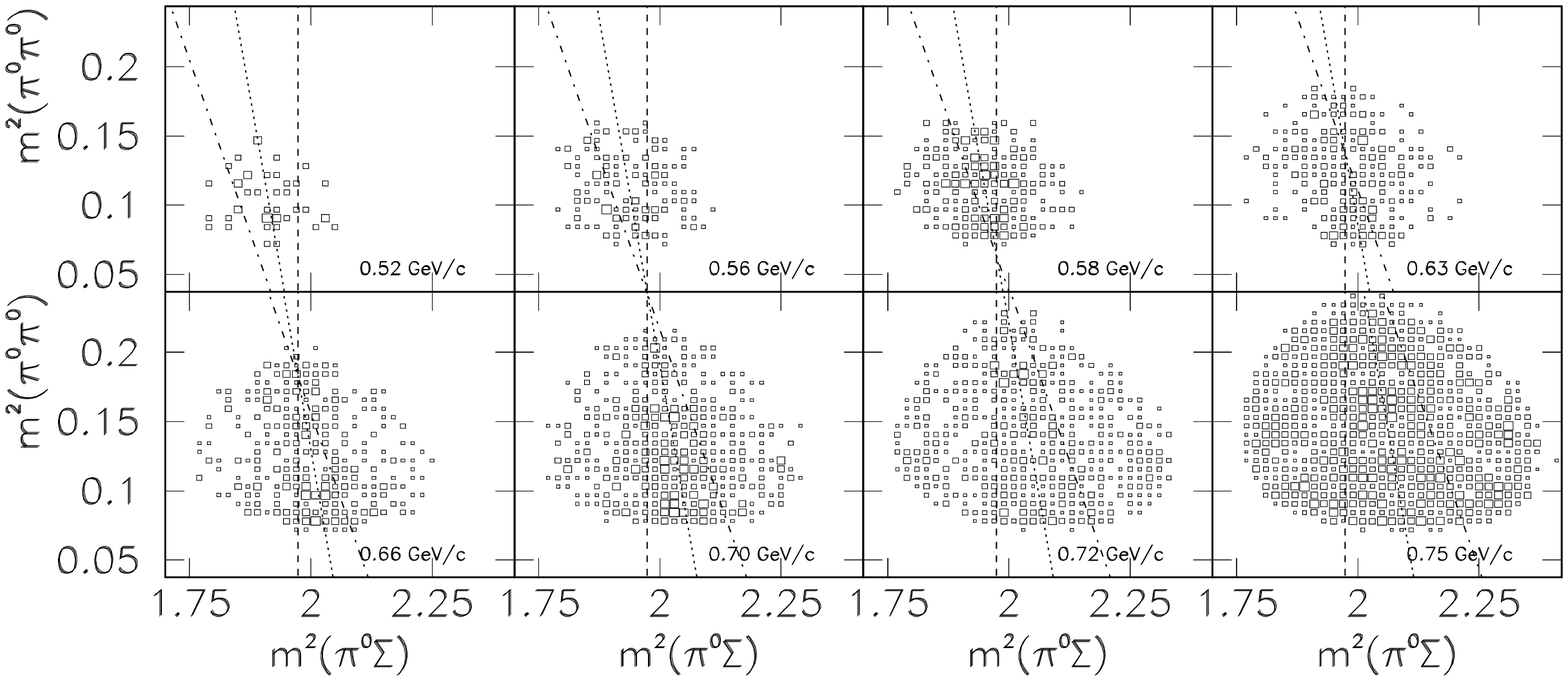}
\caption{\label{fig7}Same as Fig.~5 for the process
$K^-p\to\pi^0\pi^0\Sigma^0$. 
Dashed line: calculated position for $m^2(\pi^0\Sigma^0)=m^2(\Lambda^*)$,
with $m(\Lambda^*)=1.405$ GeV.}  
\end{figure}
Comparing Figs. 6 and 7 we see
that the  characteristic features of the $\pi^0 \pi^0 \Lambda$ and
$\pi^0 \pi^0 \Sigma^0$ final  state density distributions are
very different.  This can be expressed succinctly using the projection
of the DP on the $m^2(\pi^0\pi^0)$ axis.  This is  exhibited in
Fig.~8 for four $K^-$ beam momenta.  
\begin{figure}
\includegraphics[width=\textwidth]{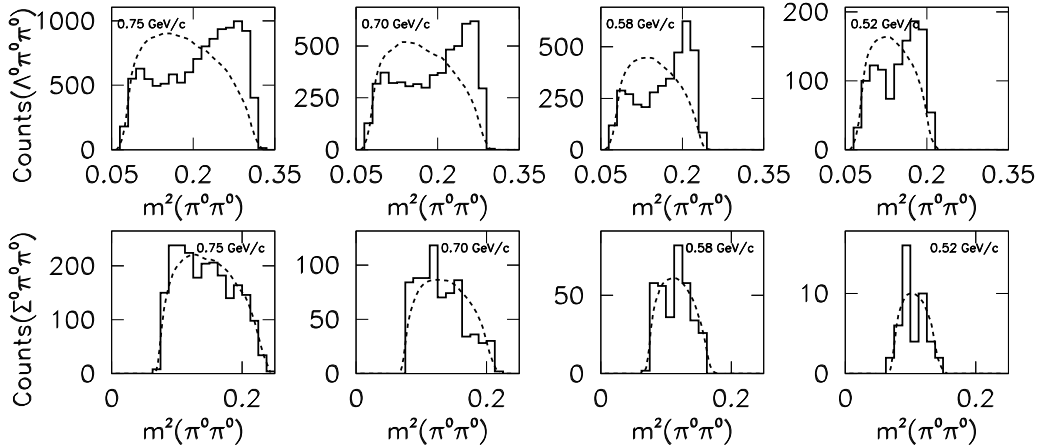}
\caption{\label{fig8}Projection of the content of the Dalitz
plots of Figs.~6 and 7 onto the $m^2(\pi^0\pi^0)$ axis.  The dashed line
is phase space.
Top row: $K^-p\to\pi^0\pi^0\Lambda$.
Bottom row: $K^-p\to\pi^0\pi^0\Sigma^0$.}  
\end{figure}
Added to the figure for 
comparison are the phase space distributions.  Note that the projection on
the $m^2(\pi^0\pi^0)$  axis is not affected by the double entry of each
event on account of the two identical $\pi^0$'s.  The projection plots
for $\pi^0\pi^0\Lambda$ have a substantial excess of events over phase space
in the upper range of $m^2(\pi^0\pi^0)$, while for $\pi^0\pi^0\Sigma^0$
a small excess occurs in the lower range.  This excludes a sizable
$f_0$ contribution for $m(f_0) \leq 450$ MeV  to $2\pi^0 $ production.
Furthermore, applying again flavor symmetry, we conclude that the
$f_0$ contribution to $K^-p\to\pi^0\pi^0\Lambda$ and
$\pi^-p\to\pi^0\pi^0n$ must be minor. 

Figure~9 shows the DP projection on the $m^2 (\pi Y^{\ast})$
axis. 
\begin{figure}
\includegraphics[width=\textwidth]{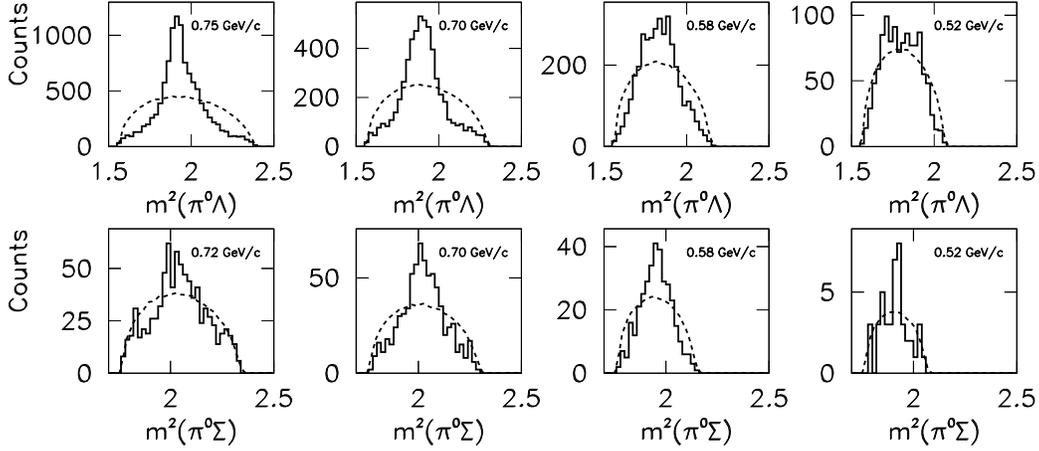}
\caption{\label{fig9}Projection of the content of the Dalitz
plots of Figs.~6 and 7 onto the horizontal axis.  The dashed line
is phase space.
Top row: $K^-p\to\pi^0\pi^0\Lambda$ projected onto the
$m^2(\pi^0\Lambda)$ axis.
Bottom row: $K^-p\to\pi^0\pi^0\Sigma^0$ projected onto the
$m^2(\pi^0\Sigma^0)$ axis.}  
\end{figure}
Recall that the spectra are distorted by the double entry of each
event on account of the two identical $\pi^0$s
in the final state.  At $p_K = 750$ MeV/$c$ the peak in $m^2 (\pi^0
\Lambda)$ is least distorted by the $2 \pi^0$ ambiguity, see Fig.~6.  The
position and width of this peak agree within error with the mass and
width of the $\Sigma^*(1385)$ resonance, this reflects $\Sigma^*$ dominance
in $K^-p\to\pi^0\pi^0\Lambda$.  For lower incident $p_K$ the peak
in $m^2(\pi^0\Lambda)$ is broadened and is shifted to lower mass as
expected, again see Fig.~6. 

The large peak in $m^2(\pi^0\Sigma^0)$ centered around (1.4
GeV)$^2$ in the bottom row of Fig.~9 reflects the dominance of the
$\Lambda^*(1405)$ resonance in $K^-p\to\pi^0\pi^0\Sigma^0$.  The
threshold for the production of this hyperon 
resonance is $p_K = 445$ MeV/$c$.  At our lowest $p_K$ of 520 MeV/$c$ we
are close to threshold of the $\Lambda (1405)$ and expect a small
cross section.  This explains the smallness of
$\sigma_{tot}(K^-p\to\pi^0\pi^0\Sigma^0)$ at this energy, see Figs.~3
and 4.  The 
decay $\Lambda(1405)\frac{1}{2}^-$ into
$\pi^0\Sigma^0(1193)\frac{1}{2}^+$ is an $\ell=0$ transition while
$\Sigma^0(1385)\frac{3}{2}^+$ decaying to
$\pi^0\Lambda(1116)\frac{1}{2}^+$ 
is $\ell = 1.$  We speculate that this is at least in part responsible
for the near uniform density of the $\Lambda(1405)$ band in
the $\pi^0 \pi^0 \Sigma^0$  DP of Fig.~7 while there is a
depletion (the island-reef structure) in the $\pi^0\pi^0\Lambda$ DP of
Fig.~6. 

The $m^2(\pi^0\Sigma^0)$ projection at $p_K$\ = 720 MeV shows
not only the dominant peak due to the $\Lambda (1405)$ in the final
state but there are two other small enhancements.  One is at
$m^2(\pi^0\Sigma^0)=(1.5\,\mathrm{GeV})^2$; it is due to the
$\Lambda$(1520) $\frac{3}{2}^-$ 
resonance.  The other enhancement  occurs at
$m^2(\pi^0\Sigma^0)=(1.36\,\mathrm{GeV})^2$; 
this is the predicted value for the reflection of the $\Lambda(1520)$.
The $\Lambda(1520)$ threshold is at $p_K = 704$ MeV/$c$. The onset of
the $\Lambda(1520)$ production explains the rise in
$\sigma_{tot} (K^- p \to \pi^0 \pi^0 \Sigma^0)$ at the two
highest $p_K$'s, see Figs.~3 and 4.

We conclude that $K^- p \to \pi^0 \pi^0 \Sigma^0$ is dominated
by $\Lambda^{\ast}$ production, there is no direct evidence for $f_0$
production for $m(f_0) \leq$ 450 MeV.  From the Dalitz plot
distribution we estimate that $f_0$ production is far less than 50\%.
Using the total cross sections and applying flavor symmetry
this implies that $f_0$ production in $K^- p \to \pi^0 \pi^0 \Lambda$
as well as in $\pi^- p \to \pi^0 \pi^0 n$ is less than 10\%.  More precise
values require more extensive analysis which is under way.

\section{Summary and Conclusions}
The reaction $K^- p \to \pi^0 \pi^0 \Sigma^0$ is dominated by
$\Lambda^{\ast}$ production, specifically the $\Lambda (1405)
\frac{1}{2}^-$ and for high $p_K$ also the $\Lambda (1520)
\frac{3}{2}^-$ resonance. The $m^2(\pi^0\pi^0)$ spectrum peaks at
\textbf{low} $m(\pi^0\pi^0)$, see Fig.~8.  The shape is ``opposite" to the
spectrum expected if the $f_0$ (or ``$\sigma$'') plays a significant
role, that would result in a peak of the DP projection at high 
$m(\pi^0\pi^0)$.  A very conservative upper limit for a possible direct
contribution of a $f_0$ with $m(f_0)\leq450$ MeV in $K^- p \to
\pi^0 \pi^0 \Sigma^0$ is half, but likely it is much smaller.
The comparison of the measured total cross sections,
$\sigma_{tot}(K^-p\to\pi^0\pi^0\Lambda) =
(5\pm1)\sigma_{tot}(K^-p\to\pi^0\pi^0\Sigma^0) \simeq 
\frac{1}{2}\sigma_{tot}(\pi^-p\to\pi^0\pi^0n)$ together with flavor
symmetry places an upper limit of 10\% 
on the $f_0$ contribution to $K^-p\to\pi^0\pi^0\Lambda$ and to
$\pi^-p\to\pi^0\pi^0n$ for $m(f_0)\lsim550$ MeV.  This is
consistent with the DP distributions for these two
reactions.  The $\pi^0 \pi^0n$ and  $\pi^0 \pi^0 \Lambda$ DP's have
very non-uniform density distributions which we have described as an
``island'' and a ``reef''.  They are due to the
preponderance of the $\Sigma(1385)\frac{3}{2}^+$ in the $\pi^0 \pi^0
\Lambda$ final state and the $\Delta(1232)\frac{3}{2}^+$ in $\pi^0
\pi^0 n$.  The DP's show that the distribution of events in the
$m^2(\pi^0\pi^0)$ spectra are not  due to the presence of an $f_0$ but due
to the 
formation of either the $\Sigma (1385)$ or the $\Delta (1232)$.

Our results on $2 \pi^0$ production on hydrogen imply that the change
in the shape of the $m^2(\pi^0\pi^0)$ distributions in $2 \pi^0$
production reported for some complex nuclei are not due in first
order to a proposed modification of the $f_0$ of a mass of about 
550 MeV in the nuclear medium.  They are more likely the result of
final state interactions of the pions and the recoil baryon states.

The remarkable similarity in the density distribution of the $\pi^0
\pi^0 n$ and $\pi^0 \pi^0 \Lambda$ final states is a convincing proof
of the  applicability of flavor symmetry to appropriate three-body
final state reactions, as is the relation
\begin{equation}
\sigma_{tot}(\pi^-p\to\pi^0\pi^0n) = 
   (2\pm0.5)\sigma_{tot}(K^-p\to\pi^0\pi^0n).
\end{equation}
Flavor symmetry does not apply to the $\pi^0\pi^0\Lambda$ and
$\pi^0\pi^0\Sigma^0$ final states so we are not surprised that they
have quite different DPs and unequal cross sections.

\end{document}